\documentclass[aps,prb,twocolumn,superscriptaddress,nofootinbib]{revtex4-2}

\usepackage{graphicx}
\usepackage{multirow}
\usepackage{amsmath}
\usepackage{placeins}
\usepackage{hyperref}

\begin{document}

\title{Domain wall motion in ferromagnetic nanowires driven by a localized Gaussian thermal gradient}

\author{M. A. Jafar Pikul}
\email{jafar.pikul@uri.edu}
\affiliation{Department of Physics, University of Rhode Island, Kingston, RI 02881, USA}
\affiliation{Physics Discipline, Khulna University, Khulna 9208, Bangladesh}

\author{M. A. S. Akanda}
\affiliation{Department of Physics and Astronomy, University of Nebraska-Lincoln, Lincoln, NE 68588, USA}

\author{M. T. Islam}
\email{torikul@phy.ku.ac.bd}
\affiliation{Physics Discipline, Khulna University, Khulna 9208, Bangladesh}

\date{\today}

\begin{abstract}
We investigate magnetic domain wall (DW) dynamics in a uniaxial ferromagnetic nanowire under the localized Gaussian temperature profile of a laser spot using the stochastic Landau-Lifshitz-Gilbert equation. The DW velocity increases linearly with peak laser temperature and decreases with increasing laser to DW distance. The velocity varies nonlinearly with Gilbert damping because damping strengthens thermal magnon excitation but shortens the magnon propagation length. The DW initially lies away from the laser-heated region, so the temperature gradient at its position is effectively zero and the entropic torque is negligible. The DW motion is therefore mainly driven by magnonic spin-transfer torque. We analyze laser temperature, laser to DW distance, damping, uniaxial anisotropy, and laser width. The analysis shows that laser width and laser to DW distance independently control the DW response. These findings may clarify the mechanism of localized thermally driven DW motion and guide thermal control strategies in spintronic racetrack-memory devices.
\end{abstract}

\maketitle


\maketitle

\section{Introduction}
Magnetic domain walls (DWs) are thin transition regions that separate oppositely magnetized domains in ferromagnetic nanowires and are promising
candidates for data storage and information processing in spintronic devices ~\cite{kumar2022domain}. In
racetrack memory, binary data are encoded as a sequence of DWs that shift
along a nanowire and are detected at fixed read
positions~\cite{Parkin2008, Hayashi2008, blasing2020magnetic}. DWs can also be used to perform logic operations~\cite{Allwood2005, venkat2024magnetic}. For these applications, DWs must move reliably, quickly, and with low energy consumption~\cite{Thiaville2005}. Several methods have been developed to drive DW motion, including applied magnetic fields, which expand domains aligned with the field but cannot move multiple DWs simultaneously in the same direction along a nanowire~\cite{Schryer1974, Beach2005, wang2009magnetic, wang2009high, wang2013}. Microwave excitation can also drive DWs, but its effectiveness strongly depends on matching the excitation frequency to the DW resonance frequency~\cite{yan2009domain, han2009magnetic}. Spin-polarized electric currents can drive DWs through spin-transfer torque, providing flexible control that is compatible with racetrack operation~\cite{Thiaville2005, Zhang2004, yamaguchi2004real, hayashi2006influence}. However, this
approach typically requires very large current densities, and resulting Joule heating can damage the
device~\cite{Tatara2004, Tatara2008, Barnes2005, yamaguchi2012temperature}.
These limitations motivate the search for cleaner and more energy-efficient
driving mechanisms. Spin caloritronics offers a promising
alternative~\cite{Bauer2012}, and in this process, a temperature gradient applied to the nanowire excites thermal magnons. These
magnons carry spin angular momentum from the hot region to the cold
region~\cite{Uchida2008, Uchida2010, Xiao2010}. When the magnons reach the DW, they transfer angular momentum to the DW via
magnonic spin-transfer torque ($\mu$STT), driving it toward the hotter
region~\cite{Yan2011, Wang2012, Kovalev2012, Ritzmann2014, Kong2013,
Chico2014}. The entropic torque~(ET) also pushes the DW
toward higher temperatures, and it arises from thermodynamic free-energy gradients
and acts only where the local temperature gradient is
nonzero~\cite{Schlickeiser2014, Wang2014, Yan2015}. The relative importance of $\mu$STT and ET depends on the material and the
shape of the temperature profile~\cite{Moretti2017, Selzer2016, Donges2020,
islam2019t}. For a uniform linear temperature gradient in Permalloy, micromagnetic simulations confirmed that $\mu$STT dominates, and the DW velocity scales linearly with the gradient magnitude~\cite{islam2019t}. Subsequent studies showed that thermally excited spin waves carrying angular momentum, often referred to as spin Seebeck waves, are attenuated by Gilbert damping ($\alpha$), which can explain the non-monotonic dependence of DW velocity on $\alpha$~\cite{akanda2023role}. The role of wire cross-section geometry has also been studied, showing that effective shape anisotropy can modify the DW speed~\cite{islam2023role2}. For a localized Gaussian temperature profile produced by a focused laser spot, the DW dynamics depend strongly on the DW position relative to the heat source. Previous stochastic Landau-Lifshitz-Bloch (LLB) simulations of Permalloy nanostrips used a broad laser profile, $\sigma_L=200$~nm, and found that both ET and $\mu$STT contribute, with ET dominating when the DW experiences a nonzero local temperature gradient~\cite{Moretti2017}. The situation changes when the laser spot is much narrower than the magnon propagation length and the DW is placed far from the laser. In this limit, the local temperature gradient at the DW position is nearly zero,
so ET is negligible, and the DW motion is dominated by $\mu$STT. This strongly
localized regime is relevant to near-field magneto-caloritronic
experiments~\cite{Pfitzner2018} and tightly focused ultrafast laser
heating~\cite{OCallahan2016}. Despite this relevance, the independent effects of laser width $\sigma_L$ and laser to DW distance $d$ on DW dynamics remain unresolved.

In this work, we investigate DW dynamics in a nanowire driven by a localized
Gaussian temperature profile generated by a laser spot. We solve the stochastic Landau-Lifshitz-Gilbert (sLLG) equation for a Permalloy nanowire under a localized laser profile with $\sigma_L=15$~nm. The micromagnetic simulations are performed below the Curie temperature, where the sLLG description remains appropriate, and yield results consistent with the Landau-Lifshitz-Bloch (LLB) formulation~\cite{Moretti2017}. By independently varying $\sigma_L$ and $d$, we separate their effects on DW motion. We also perform a systematic damping study in the sub-Walker regime, where DWs move without precessional instability. The resulting non-monotonic velocity dependence provides a direct signature of the magnonic spin-transfer torque mechanism. Together with studies of laser temperature, separation, and anisotropy, this work provides a parameter map for thermally driven DW dynamics in the strongly localized Gaussian region.

\section{Model and simulation method}

\begin{figure}
  \centering
  \includegraphics[width=0.8\columnwidth]{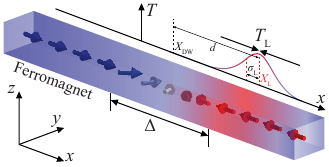}
  \caption{%
    Schematic of the Permalloy nanowire with length $1024$~nm and cross-section
    $8\times8$~nm$^2$, containing a head-to-head domain wall (DW) at the center.
    The red curve represents the localized Gaussian temperature profile, which
    peaks at the laser position $X_L$. The laser spot is located at a distance
    $d = N\sigma_L$ to the right of the initial DW position $X_{\rm DW}$, where
    $\sigma_L$ is the laser width. Blue and red coloring indicate the colder and
    hotter regions, respectively. The DW width $\Delta$ and the laser to DW
    distance $d$ are also indicated.}
  \label{Fig1}
\end{figure}

We simulate a uniaxial ferromagnetic nanowire of length $L_x = 1024$~nm with
a square cross-section $L_y \times L_z = 8\times8$~nm$^2$ where the easy axis is
along $\hat{x}$. A head-to-head DW is placed at the wire center,
$X_{\rm DW} = 512$~nm, as shown in Fig.~\ref{Fig1}. A focused laser spot deposits heat that follows a Gaussian profile along the
wire~\cite{Moretti2017},
\begin{equation}
T(x) = T_L\,\exp\!\left[-\frac{(x - X_L)^2}{2\sigma_L^2}\right],
\label{eq:Tprofile}
\end{equation}
where $T_L$ is the peak temperature at laser position $X_L$ and $\sigma_L$
is the laser width. The temperature drops to $T_L/e$ at a distance $\sqrt{2}\,\sigma_L \approx 21$~nm
from the laser center, defining the effective width of the heated region. The laser is placed
to the right of the DW at $X_L = X_{\rm DW} + d$, where $d = N\sigma_L$ and
$N$ is a positive integer. Here we used $\sigma_L = 15$~nm, $N = 7$ ($d = 105$~nm), and
$T_L = 300$~K. The normalized magnetization $\mathbf{m} = \mathbf{M}/M_s$ evolves under the
stochastic Landau-Lifshitz-Gilbert (sLLG)
equation~\cite{Landau1935, gilbert2004, brown1963},
\begin{equation}
\frac{d\mathbf{m}}{dt} =
  -\gamma\,\mathbf{m}\times\!\left(\mathbf{h}_{\rm eff}
  + \mathbf{h}_{\rm th}\right)
  + \alpha\,\mathbf{m}\times\frac{d\mathbf{m}}{dt},
\label{eq:sLLG}
\end{equation}
where $M_s$ is the saturation magnetization, $\gamma$ is the gyromagnetic
ratio, and $\alpha$ is the Gilbert damping constant. The deterministic
effective field is,
\begin{equation}
\mathbf{h}_{\rm eff}
= \frac{2A_{\rm ex}}{\mu_0 M_s}
  \sum_{\sigma=1}^{3}\frac{\partial^2 \mathbf{m}}{\partial x_\sigma^2}
+ \frac{2K_u}{\mu_0 M_s}\,m_x\,\hat{\mathbf{x}}
+ \mathbf{h}_{\rm dipole},
\label{eq:heff}
\end{equation}
where $A_{\rm ex}$ is the exchange stiffness, $K_u$ is the uniaxial anisotropy
constant, and $\mathbf{h}_{\rm dipole}$ is the dipolar field.
The stochastic thermal field $\mathbf{h}_{\rm th}$ satisfies the
fluctuation dissipation theorem~\cite{hinzke2008},
\begin{equation}
\begin{aligned}
\langle h_{{\rm th},ip}(t)\rangle &= 0,\\
\bigl\langle h_{{\rm th},ip}(t)\,h_{{\rm th},jq}(t+\Delta t)\bigr\rangle
  &= \frac{2k_B T_i \alpha_i}{\gamma\mu_0 M_s a^3}\,
     \delta_{ij}\,\delta_{pq}\,\delta(\Delta t),
\end{aligned}
\label{eq:fdt}
\end{equation}
where $i$ and $j$ label micromagnetic cells, $p$ and $q$ denote Cartesian
components, $T_i$ and $\alpha_i$ are the local temperature and damping at
cell~$i$, $a$ is the cell size, and $k_B$ is the Boltzmann constant. The discretized
thermal field at each cell is~\cite{martinez2007thermal},
\begin{equation}
\begin{aligned}
h_{{\rm th},ip}
= \eta\,\sqrt{\frac{2\alpha_i k_B T_i}{\gamma\mu_0 M_s a^3\,\Delta t}},
\label{eq:hth}
\end{aligned}
\end{equation}
where $\eta$ is a random number drawn from a standard normal distribution, and $\Delta t$ is the integration time step.

We solve the sLLG equation, Eq.~\eqref{eq:sLLG}, numerically in
MuMax3~\cite{mumax} using the adaptive Heun method. The sLLG framework is
valid here because all simulated temperatures are below the Curie
temperature of Permalloy, $T_c = 850$~K~\cite{Moretti2017}. The simulated material is Permalloy
(Ni$_{80}$Fe$_{20}$)~\cite{islam2019t}, with saturation magnetization
$M_s = 8\times10^5$~A m$^{-1}$, exchange stiffness
$A_{\rm ex} = 13\times10^{-12}$~J m$^{-1}$, and uniaxial anisotropy
$K_u = 2\times10^5$~J m$^{-3}$. The simulation grid is discretized into $512\times4\times4$ cells, with a cell size of $2\times2\times2$~nm$^3$. This gives a total simulation size of
$1024\times8\times8$~nm$^3$. The integration time
step is $\Delta t = 10^{-14}$~s~\cite{akanda2023role, islam2023role2, islam2021thermally}. We remove the magnetostatic surface charges at the wire boundaries to prevent
the DW from being artificially attracted toward the wire ends~\cite{martinez2007thermal}. For each parameter set, results are
averaged over 15 independent simulations with different random seeds. Error
bars represent one standard deviation.

\section{Numerical results}
\begin{figure}
  \centering
  \includegraphics[width=1\columnwidth]{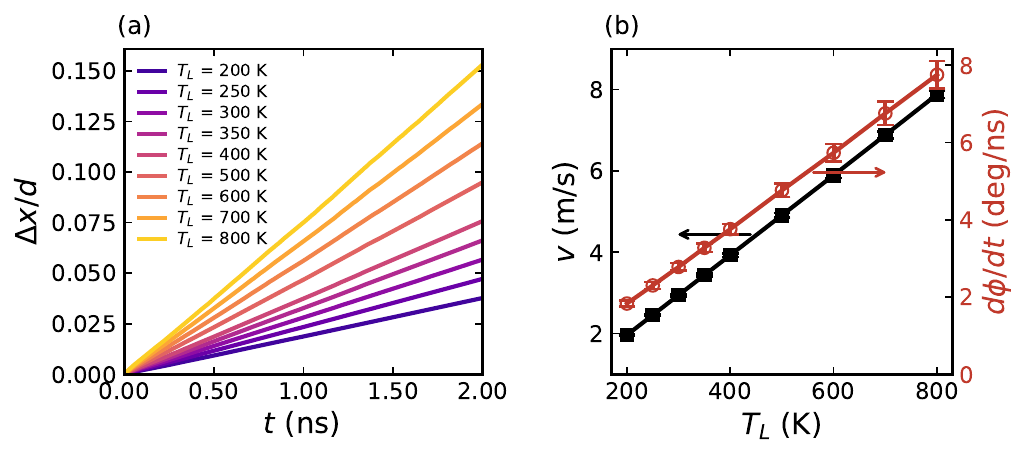}
  \caption{%
    Effect of laser temperature on DW dynamics for fixed
    $N = 7$ ($d = 105$~nm), $\sigma_L = 15$~nm,
    $\alpha = 0.001$, and $K_u = 2\times10^5$~J m$^{-3}$.
    (a)~Normalized DW displacement $\Delta x/d$ as a function of time for
    $T_L =$ 200, 250, 300, 350, 400, 500, 600, 700, and 800~K. The DW moves
    toward the hotter region for all temperatures, and the displacement increases
    with $T_L$.
    (b)~Translational velocity $v$ (filled black squares) and angular velocity
    $\mathrm{d}\phi/\mathrm{d}t$ (open red circles) as functions of $T_L$.
    Both quantities increase linearly with laser temperature. Error bars
    represent one standard deviation over 15 independent simulations.}
  \label{fig:TL}
\end{figure}

Figure~\ref{fig:TL} shows the DW response as the peak laser temperature $T_L$ is varied from 200 to 800~K (all values below $T_c=850$~K), with $N=7$ ($d=105$~nm), $\sigma_L=15$~nm, and $\alpha=0.001$ fixed. The normalized DW displacement $\Delta x/d$ as
a function of time is presented in Fig.~\ref{fig:TL}(a). In every case, the DW moves toward the hotter region, as expected for
magnonic spin-transfer torque in which magnons propagate from hot to cold and
transfer angular momentum to the DW upon arrival~\cite{Yan2011, Wang2012, islam2019t}. The DW displacement increases with $T_L$, since a higher laser
temperature produces a more intense thermal field that excites more magnons
and exerts a stronger torque on the DW. For $T_L \geq 400$~K, the DW accelerates as it approaches the laser-heated
region, where thermal magnon excitation is strongest. Since the DW is initially
located at a distance $d = 7\sigma_L$ from the laser position, the localized
temperature gradient at the DW position is effectively zero. Thus, the entropic
torque is negligible in this regime, and the observed motion is driven primarily
by magnonic spin-transfer torque from magnons propagating from the laser spot
to the DW~\cite{Yan2011, Moretti2017}. Figure~\ref{fig:TL}(b) shows that both the translational velocity $v$ and
angular velocity $\mathrm{d}\phi/\mathrm{d}t$ increase linearly with
$T_L$. This trend is consistent with a thermal mobility picture in which the
DW velocity scales with the characteristic strength of the temperature profile,
which is proportional to $T_L/\sigma_L$~\cite{islam2019t, Schlickeiser2014,
Wang2014}. From the slope of the linear fit, we estimate the thermal mobility
coefficient as $\mu_T = v\,\sigma_L/T_L \approx 1.48\times10^{-10}$~m$^2$ s$^{-1}$
K$^{-1}$. The angular velocity increases with $T_L$ because stronger thermal excitation
enhances the magnonic spin current, which drives both DW translation and
precessional motion~\cite{islam2019t, Yan2011}.

\begin{figure}
  \centering
  \includegraphics[width=1\columnwidth]{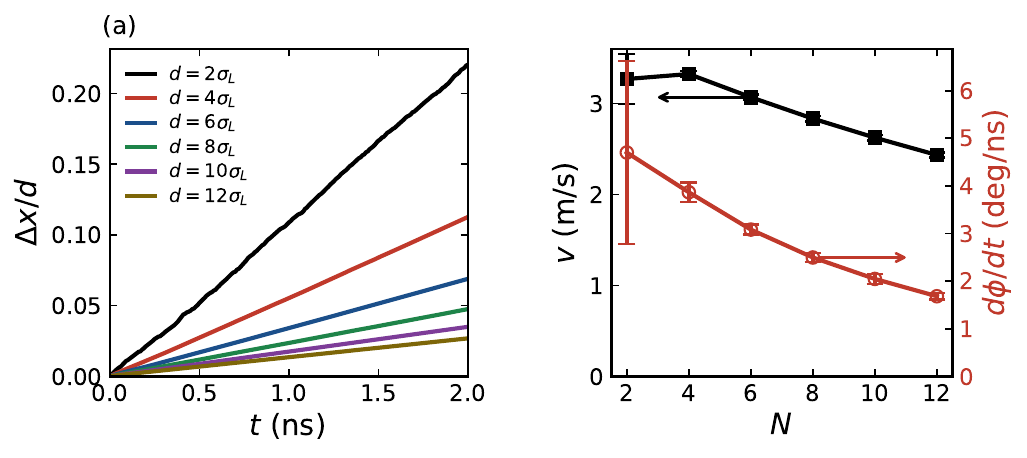}
  \caption{%
    Dependence of DW dynamics on laser to DW distance for fixed
    $T_L = 300$~K, $\sigma_L = 15$~nm, and $\alpha = 0.001$.
    (a)~Normalized displacement $\Delta x/d$ versus time for
    $N = 2, 4, 6, 8, 10, 12$, where $d = N\sigma_L$.
    (b)~Translational velocity $v$ (filled black squares) and angular velocity
    $\mathrm{d}\phi/\mathrm{d}t$ (open red circles) versus $N$. Both quantities
    decrease with increasing $N$ because the magnon current weakens as it travels
    farther from the laser spot to the DW.}
  \label{fig:N}
\end{figure}

We fix $T_L = 300$~K, $\sigma_L = 15$~nm, and $\alpha = 0.001$, and vary the
separation parameter $N \in \{2, 4, 6, 8, 10, 12\}$. This changes the physical
laser to DW distance according to $d = N\sigma_L$. Figure~\ref{fig:N}(a) shows
the normalized displacement $\Delta x/d$ as a function of time for each value
of $N$. The displacement decreases systematically as $N$ increases. At larger
separations, the magnon current must travel farther before reaching the DW,
and damping reduces its intensity during propagation. This attenuation is
consistent with the exponential decay of spin-current density with distance
predicted for magnon transport in a damped medium~\cite{Ritzmann2014,
Kovalev2012, Moretti2017}. Figure~\ref{fig:N}(b) shows the translational velocity $v$ and angular velocity
$\mathrm{d}\phi/\mathrm{d}t$ as functions of $N$. Both quantities decrease with
increasing $N$. The angular velocity decreases more sharply than the
translational velocity at large $N$, indicating that the precessional response
is more sensitive to the attenuation of the magnon spin current at the DW
location~\cite{islam2019t, akanda2023role}.

\begin{figure}
  \centering
  \includegraphics[width=0.6\columnwidth]{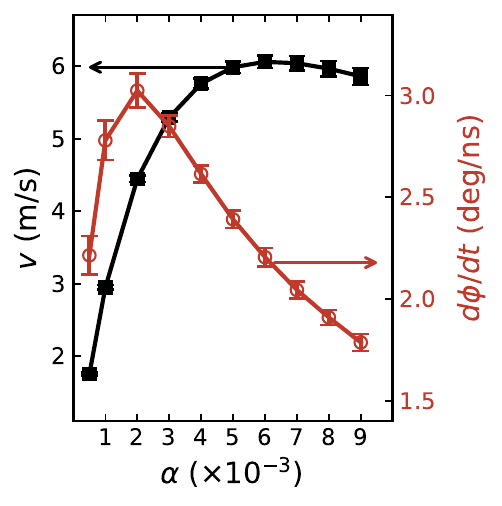}
 \caption{%
Damping dependence of DW dynamics in the sub-Walker regime for fixed
$T_L = 300$~K, $N = 7$, $\sigma_L = 15$~nm, and
$K_u = 2\times10^5$~J m$^{-3}$.
Translational velocity $v$ (filled black squares) and angular velocity
$\mathrm{d}\phi/\mathrm{d}t$ (open red circles) are plotted as functions of
$\alpha$ ($\times 10^{-3}$). The velocity reaches a broad maximum near
$\alpha \approx 0.002$--$0.003$ and then levels off, while
$\mathrm{d}\phi/\mathrm{d}t$ decreases after reaching its maximum near
$\alpha \approx 0.002$. This behavior reflects the competition between
magnon generation and magnon propagation.}
  \label{fig:alpha}
\end{figure}

The dependence of $v$ and $\mathrm{d}\phi/\mathrm{d}t$ on Gilbert damping
is observed in the sub-Walker regime~\cite{Markus2022}, with all other parameters fixed at
$T_L = 300$~K, $N = 7$, $\sigma_L = 15$~nm, and
$K_u = 2\times10^5$~J m$^{-3}$, as shown in Fig.~\ref{fig:alpha}.
The translational velocity $v$ increases rapidly at low damping, reaches a
broad maximum near $\alpha \approx 0.002$--$0.003$, and then levels off at
about $6$~m s$^{-1}$. The angular velocity
$\mathrm{d}\phi/\mathrm{d}t$ reaches its maximum near
$\alpha \approx 0.002$ and then decreases continuously.
Both trends are consistent with $\mu$STT as the driving
mechanism~\cite{islam2019t, akanda2023role}. The non-monotonic behavior of
$v$ arises from two competing effects. First, the stochastic thermal field
amplitude scales as $\sqrt{\alpha}$~\cite{brown1963}, so fewer magnons are
excited at small $\alpha$. Second, the magnon propagation length scales as
$\alpha^{-1/2}$~\cite{Ritzmann2014, Yan2011}, so magnons are attenuated more
strongly at larger $\alpha$. At intermediate damping, magnon generation and
propagation are well balanced, producing the velocity maximum. The continuous
decrease of $\mathrm{d}\phi/\mathrm{d}t$ with increasing $\alpha$ reflects
stronger dissipation of spin precession, reducing the net angular momentum
delivered to the DW per unit time~\cite{islam2019t, akanda2023role}. This
non-monotonic damping dependence is absent in field-driven
models~\cite{Schryer1974}, making the observed peak in $v$ a clear
signature of $\mu$STT.

\begin{figure}
  \centering
  \includegraphics[width=\columnwidth]{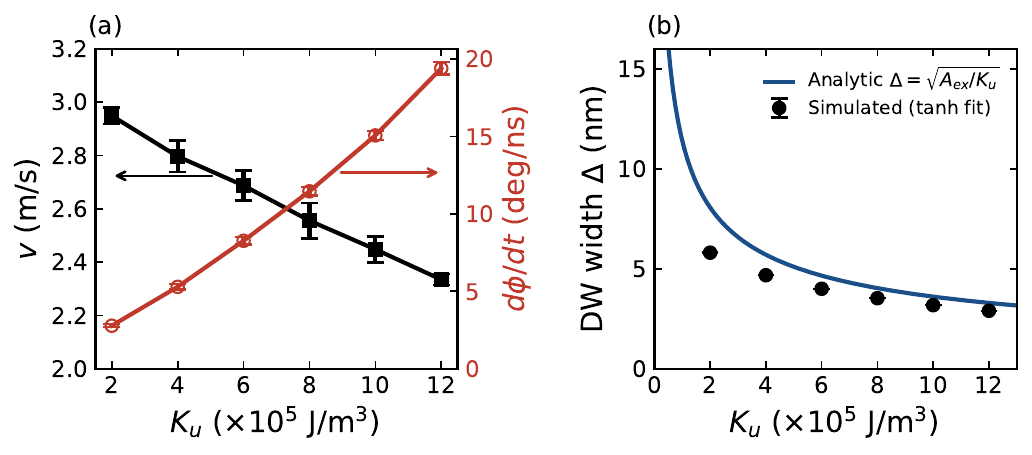}
    \caption{%
    Uniaxial anisotropy study for fixed $T_L = 300$~K, $N = 7$,
    $\sigma_L = 15$~nm, and $\alpha = 0.001$.
    (a)~Translational velocity $v$ (filled black squares) and angular velocity
    $\mathrm{d}\phi/\mathrm{d}t$ (open red circles) plotted against $K_u$.
    As $K_u$ increases, $v$ decreases while $\mathrm{d}\phi/\mathrm{d}t$
    increases. (b)~DW width $\Delta$ plotted against $K_u$. Filled black circles represent
    values extracted from $\tanh$ fits to the simulated magnetization profiles,
    and the blue curve represents the analytic  prediction,
    $\Delta = \sqrt{A_{\rm ex}/K_u}$. The decreasing width confirms that the DW
    narrows with increasing uniaxial anisotropy.}
  \label{fig:Ku}
\end{figure}

The role of uniaxial anisotropy $K_u$ is studied with
$T_L = 300$~K, $N = 7$, $\sigma_L = 15$~nm, and
$\alpha = 0.001$, as shown in Fig.~\ref{fig:Ku}.
Figure~\ref{fig:Ku}(a) shows that the translational velocity $v$ decreases
as $K_u$ is increased from $2\times10^5$~J m$^{-3}$ to
$12\times10^5$~J m$^{-3}$, whereas the angular velocity
$\mathrm{d}\phi/\mathrm{d}t$ increases over the same range. This indicates
that increasing anisotropy affects the translational and precessional
responses differently. Figure~\ref{fig:Ku}(b) shows the DW width $\Delta$ as a function of $K_u$.
We extract $\Delta$ by fitting the simulated magnetization profile $m_x(x)$
to the form $m_x = -\tanh[(x - X_{\rm DW})/\Delta]$. The DW width decreases
with increasing $K_u$, consistent with the analytic scaling
$\Delta = \sqrt{A_{\rm ex}/K_u}$~\cite{Nakatani2003, Thiaville2005}. This
confirms that the DW becomes narrower as the uniaxial anisotropy increases,
in agreement with the standard one-dimensional DW model~\cite{Schryer1974,
Nakatani2003}. Here, $K_u$ represents crystalline anisotropy, which is distinct from the
effective shape anisotropy produced by the dipolar field of the wire
cross-section~\cite{islam2023role2}. Therefore, these simulations show the
influence of $K_u$ without changing the wire geometry, separating crystalline
anisotropy from shape anisotropy effects.

\begin{figure}
  \centering
  \includegraphics[width=\columnwidth]{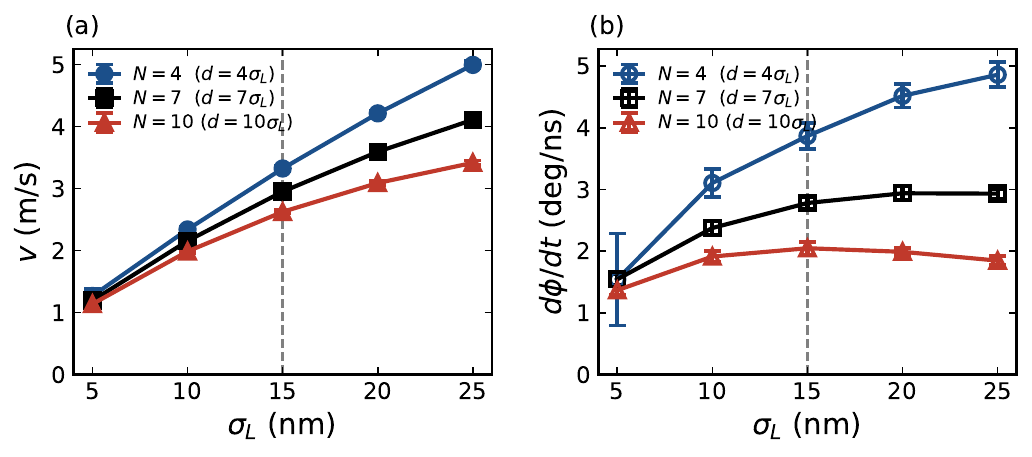}
\caption{%
Laser width dependence of DW dynamics for $N = 4$, $7$, and $10$, with
$T_L = 300$~K, $\alpha = 0.001$, and
$K_u = 2\times10^5$~J m$^{-3}$ fixed.
(a)~Translational velocity $v$ as a function of $\sigma_L$ (filled markers).
(b)~Angular velocity $\mathrm{d}\phi/\mathrm{d}t$ as a function of
$\sigma_L$ (open markers).
The dashed vertical line marks the default laser width, $\sigma_L = 15$~nm. The separated curves indicate that $\sigma_L$ and $N$ act as independent control parameters for DW motion.}
  \label{fig:sigma}
\end{figure}

The laser width $\sigma_L$ is varied for three fixed values of the separation
parameter, $N \in \{4, 7, 10\}$, as shown in Fig.~\ref{fig:sigma}. The other
parameters are fixed at $T_L = 300$~K, $\alpha = 0.001$, and
$K_u = 2\times10^5$~J m$^{-3}$. The laser width takes the values
$\sigma_L \in \{5, 10, 15, 20, 25\}$~nm. Figure~\ref{fig:sigma}(a) shows that the translational velocity $v$ increases
with $\sigma_L$ for all three values of $N$. A wider laser spot heats a larger
region of the wire, generating more thermal magnons and producing a stronger
magnonic spin current toward the DW. As a result, the translational velocity
increases with laser width. Figure~\ref{fig:sigma}(b) shows that the angular
velocity $\mathrm{d}\phi/\mathrm{d}t$ generally increases with $\sigma_L$ for
smaller separations, while the curve for $N = 10$ reaches a maximum near
$\sigma_L = 15$~nm and then decreases slightly. The curves for $N = 4$, $7$, and $10$ remain separated in both panels,
showing that $N$ and $\sigma_L$ affect the DW dynamics independently. At any
fixed $\sigma_L$, larger $N$ gives slower DW motion, consistent with the
distance dependent attenuation discussed in Fig.~\ref{fig:N}. This behavior
differs from linear gradient studies~\cite{islam2019t, akanda2023role}, where
the DW velocity is mainly controlled by the gradient magnitude. In the
strongly localized regime, both the width of the heated region and the
laser to DW distance independently shape the DW dynamics~\cite{Moretti2017}.

\begin{figure}
  \centering
  \includegraphics[width=\columnwidth]{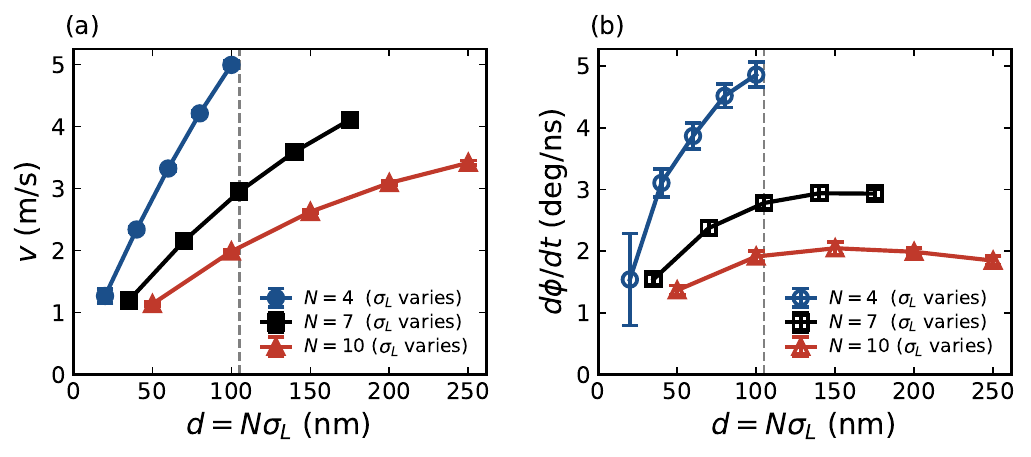}
  \caption{%
    DW dynamics plotted against the physical laser to DW distance
    $d = N\sigma_L$ for $N = 4$, $7$, and $10$, with
    $T_L = 300$~K, $\alpha = 0.001$, and
    $K_u = 2\times10^5$~J m$^{-3}$ fixed.
    (a)~Translational velocity $v$ versus $d$, shown with filled markers.
    (b)~Angular velocity $\mathrm{d}\phi/\mathrm{d}t$ versus $d$, shown with
    open markers. The dashed vertical line marks the default distance
    $d = 105$~nm. The curves do not collapse onto a single trend, confirming
    that $\sigma_L$ and the laser to DW distance act as independent control
    parameters for DW motion.}
  \label{fig:d}
\end{figure}

The independent roles of $\sigma_L$ and $N$ are tested more directly by
plotting $v$ and $\mathrm{d}\phi/\mathrm{d}t$ against the physical
laser to DW distance $d = N\sigma_L$, as shown in Fig.~\ref{fig:d}. If $d$
were the only relevant length scale, the curves for different $N$ would
collapse onto a single trend. Instead, the curves remain clearly separated,
showing that the DW dynamics cannot be determined from $d$ alone. This behavior reflects the separate roles of propagation and magnon generation.
The propagation distance $d$ determines the attenuation of the magnon current
before it reaches the DW~\cite{Ritzmann2014, Kovalev2012}, whereas the laser
width $\sigma_L$ sets the size of the heated region, the magnon generation
rate, and the spin-current profile~\cite{Moretti2017, Mochizuki2014}.
Increasing $\sigma_L$ at fixed $N$ increases $d$ while also heating a larger
region and generating more magnons. By contrast, increasing $N$ at fixed
$\sigma_L$ increases $d$ without changing the heated-region width. These two
operations are not equivalent, which explains why the curves do not collapse
onto a single function of $d$.

\section{Discussion and conclusions}

We have studied domain wall (DW) motion in a uniaxial ferromagnetic nanowire heated by a localized Gaussian laser spot using stochastic Landau-Lifshitz-Gilbert simulations. The DW is initially located away from the laser-heated region, so the temperature gradient at the DW position is effectively zero. The entropic torque is therefore negligible, and the motion is mainly driven by magnonic spin-transfer torque from thermally excited magnons. The results show that the DW velocity increases linearly with peak laser temperature and decreases with increasing laser to DW distance as the magnon current attenuates during propagation. The velocity also exhibits a non-monotonic dependence on Gilbert damping because
increasing damping strengthens thermal magnon excitation while reducing the
magnon propagation length. This damping dependence provides a clear signature of the magnonic driving mechanism. Increasing the uniaxial anisotropy narrows the DW and affects translation and precession differently, with $v$ decreasing while $\mathrm{d}\phi/\mathrm{d}t$ increases. The findings further indicate that laser width $\sigma_L$ and laser to DW distance $d$ cannot be reduced to a single control length. The laser width controls the heated-region size and magnon generation, whereas $d$ controls propagation attenuation. This explains why the velocity curves do not collapse onto a single function of $d$. For peak temperatures below 500~K, laser to DW separations of
50--150~nm, laser widths of 10--20~nm, and $\alpha \approx 0.001$--$0.003$
provide a useful operating range for thermal control of DW motion in
racetrack-memory devices.

\section*{Acknowledgments}
We thank the Institute for Theoretical and Computational Physics Research
(ITCPR), Bangladesh, for providing the platform and resources for this study. M. T. Islam acknowledges the support of the Research and Innovation Center, Khulna University (KUIRC-RPG-39/2025).

\bibliographystyle{apsrev4-2}
\bibliography{bibliography}

@article{kumar2022domain,
  title   = {Domain wall memory: Physics, materials, and devices},
  author  = {Kumar, Durgesh and Jin, Tianli and Sbiaa, Rachid and Kl{\"a}ui,
             Mathias and Bedanta, Subhankar and Fukami, Shunsuke and
             Ravelosona, Dafine and Yang, See-Hun and Liu, Xiaoxi and
             Piramanayagam, SN},
  journal = {Phys. Rep.},
  volume  = {958},
  pages   = {1--35},
  year    = {2022},
  publisher = {Elsevier},
  note    = {\url{https://doi.org/10.1016/j.physrep.2022.02.001}}
}

@article{Parkin2008,
  author  = {Parkin, Stuart S. P. and Hayashi, Masamitsu and Thomas, Luc},
  title   = {Magnetic Domain-Wall Racetrack Memory},
  journal = {Science},
  volume  = {320},
  pages   = {190--194},
  year    = {2008},
  note    = {\url{https://doi.org/10.1126/science.1145799}}
}

@article{Hayashi2008,
  author  = {Hayashi, Masamitsu and Thomas, Luc and Moriya, Rai
             and Rettner, Charles and Parkin, Stuart S. P.},
  title   = {Current-Controlled Magnetic Domain-Wall Nanowire Shift Register},
  journal = {Science},
  volume  = {320},
  pages   = {209--211},
  year    = {2008},
  note    = {\url{https://doi.org/10.1126/science.1154587}}
}

@article{blasing2020magnetic,
  title   = {Magnetic racetrack memory: from physics to the cusp of
             applications within a decade},
  author  = {Bl{\"a}sing, R. and Khan, A. A. and Filippou, P. C. and
             Garg, C. and Hameed, F. and Castrillon, J. and Parkin, S. S. P.},
  journal = {Proc. IEEE},
  volume  = {108},
  number  = {8},
  pages   = {1303--1321},
  year    = {2020},
  note    = {\url{https://doi.org/10.1109/JPROC.2020.2975719}}
}

@article{Allwood2005,
  author  = {Allwood, D. A. and Xiong, G. and Faulkner, C. C.
             and Atkinson, D. and Petit, D. and Cowburn, R. P.},
  title   = {Magnetic Domain-Wall Logic},
  journal = {Science},
  volume  = {309},
  pages   = {1688--1692},
  year    = {2005},
  note    = {\url{https://doi.org/10.1126/science.1108813}}
}

@article{venkat2024magnetic,
  title   = {Magnetic domain walls: types, processes and applications},
  author = {Venkat, G and Allwood, D A and Hayward, T J},
  journal = {J. Phys. D: Appl. Phys.},
  volume  = {57},
  pages   = {063001},
  year    = {2024},
  note    = {\url{https://doi.org/10.1088/1361-6463/ad0568}}
}

@article{Thiaville2005,
  author  = {Thiaville, Andr{\'e} and Nakatani, Yoshinobu and Miltat,
             Jacques and Suzuki, Yasuhiro},
  title   = {Micromagnetic understanding of current-driven domain wall
             motion in patterned nanowires},
  journal = {EPL},
  volume  = {69},
  pages   = {990--996},
  year    = {2005},
  note    = {\url{https://doi.org/10.1209/epl/i2004-10452-6}}
}

@article{Schryer1974,
  author  = {Schryer, N. L. and Walker, L. R.},
  title   = {The motion of 180\textdegree{} domain walls in uniform dc
             magnetic fields},
  journal = {J. Appl. Phys.},
  volume  = {45},
  pages   = {5406--5421},
  year    = {1974},
  note    = {\url{https://doi.org/10.1063/1.1663252}}
}

@article{Beach2005,
  author  = {Beach, Geoffrey S. D. and Nistor, Corneliu and Knutson, Carl
             and Tsoi, Maxim and Erskine, James L.},
  title   = {Dynamics of field-driven domain-wall propagation in
             ferromagnetic nanowires},
  journal = {Nat. Mater.},
  volume  = {4},
  pages   = {741--744},
  year    = {2005},
  note    = {\url{https://doi.org/10.1038/nmat1477}}
}

@article{wang2009magnetic,
  title   = {Magnetic field driven domain-wall propagation in magnetic
             nanowires},
  author  = {Wang, XR and Yan, P and Lu, J and He, C},
  journal = {Ann. Phys.},
  volume  = {324},
  number  = {8},
  pages   = {1815--1820},
  year    = {2009},
publisher = {Elsevier},
  note    = {\url{https://doi.org/10.1016/j.aop.2009.05.004}}
}

@article{wang2009high,
  title   = {High-field domain wall propagation velocity in magnetic
             nanowires},
  author  = {Wang, XR and Yan, P and Lu, J},
  journal = {EPL},
  volume  = {86},
  number  = {6},
  pages   = {67001},
  year    = {2009},
  note    = {\url{https://doi.org/10.1209/0295-5075/86/67001}}
}

@article{wang2013,
  author  = {Hu, B. and Wang, X. R.},
  title   = {Instability of {W}alker Propagating Domain Wall in Magnetic
             Nanowires},
  journal = {Phys. Rev. Lett.},
  volume  = {111},
  pages   = {027205},
  year    = {2013},
  note    = {\url{https://doi.org/10.1103/PhysRevLett.111.027205}}
}

@article{yan2009domain,
  title   = {Domain wall propagation due to the synchronization with
             circularly polarized microwaves},
  author  = {Yan, Peng and Wang, XR},
  journal = {Phys. Rev. B},
  volume  = {80},
  number  = {21},
  pages   = {214426},
  year    = {2009},
  publisher = {APS},
  note    = {\url{https://doi.org/10.1103/PhysRevB.80.214426}}
}

@article{han2009magnetic,
  title   = {Magnetic domain-wall motion by propagating spin waves},
  author  = {Han, Dong-Soo and Kim, Sang-Koog and Lee, Jun-Young and
             Hermsdoerfer, Sebastian J and Schultheiss, Helmut and
             Leven, Britta and Hillebrands, Burkard},
  journal = {Appl. Phys. Lett.},
  volume  = {94},
  pages = {112502},
  number  = {11},
  year    = {2009},
  publisher = {AIP Publishing},
  note    = {\url{https://doi.org/10.1063/1.3098409}}
}

@article{Zhang2004,
  author  = {Zhang, S. and Li, Z.},
  title   = {Roles of Nonequilibrium Conduction Electrons on the
             Magnetization Dynamics of Ferromagnets},
  journal = {Phys. Rev. Lett.},
  volume  = {93},
  pages   = {127204},
  year    = {2004},
  note    = {\url{https://doi.org/10.1103/PhysRevLett.93.127204}}
}

@article{yamaguchi2004real,
  title   = {Real-space observation of current-driven domain wall motion
             in submicron magnetic wires},
  author  = {Yamaguchi, Akinobu and Ono, Teruo and Nasu, Saburo and
             Miyake, Kousaku and Mibu, Ko and Shinjo, Teruya},
  journal = {Phys. Rev. Lett.},
  volume  = {92},
  number  = {7},
  pages   = {077205},
  year    = {2004},
  publisher = {APS},
  note    = {\url{https://doi.org/10.1103/PhysRevLett.92.077205}}
}

@article{hayashi2006influence,
  title   = {Influence of Current on Field-Driven Domain Wall Motion in
             Permalloy Nanowires from Time Resolved Measurements of
             Anisotropic Magnetoresistance},
  author  = {Hayashi, M and Thomas, L and Bazaliy, Ya B and Rettner, C
             and Moriya, R and Jiang, X and Parkin, SSP},
  journal = {Phys. Rev. Lett.},
  volume  = {96},
  number  = {19},
  pages   = {197207},
  year    = {2006},
  publisher = {APS},
  note    = {\url{https://doi.org/10.1103/PhysRevLett.96.197207}}
}

@article{Tatara2004,
  author  = {Tatara, Gen and Kohno, Hiroshi},
  title   = {Theory of Current-Driven Domain Wall Motion: Spin Transfer
             versus Momentum Transfer},
  journal = {Phys. Rev. Lett.},
  volume  = {92},
  pages   = {086601},
  year    = {2004},
  note    = {\url{https://doi.org/10.1103/PhysRevLett.92.086601}}
}

@article{Tatara2008,
  author  = {Tatara, Gen and Kohno, Hiroshi and Shibata, Junya},
  title   = {Microscopic approach to current-driven domain wall dynamics},
  journal = {Phys. Rep.},
  volume  = {468},
  pages   = {213--301},
  year    = {2008},
  note    = {\url{https://doi.org/10.1016/j.physrep.2008.07.003}}
}

@article{Barnes2005,
  author  = {Barnes, S. E. and Maekawa, S.},
  title   = {Current-Spin Coupling for Ferromagnetic Domain Walls
             in Fine Wires},
  journal = {Phys. Rev. Lett.},
  volume  = {95},
  pages   = {107204},
  year    = {2005},
  note    = {\url{https://doi.org/10.1103/PhysRevLett.95.107204}}
}

@article{yamaguchi2012temperature,
  title   = {Temperature estimation in a ferromagnetic {Fe-Ni} nanowire
             involving a current-driven domain wall motion},
  author  = {Yamaguchi, Akinobu and Hirohata, Atsufumi and Ono, Teruo
             and Miyajima, Hideki},
  journal = {J. Phys.: Condens. Matter},
  volume  = {24},
  number  = {2},
  pages   = {024201},
  year    = {2012},
  publisher = {IOP Publishing},
  note    = {\url{https://doi.org/10.1088/0953-8984/24/2/024201}}
}

@article{Bauer2012,
  author  = {Bauer, Gerrit E. W. and Saitoh, Eiji and van Wees, Bart J.},
  title   = {Spin caloritronics},
  journal = {Nat. Mater.},
  volume  = {11},
  pages   = {391--399},
  year    = {2012},
  note    = {\url{https://doi.org/10.1038/nmat3301}}
}

@article{Uchida2008,
  author  = {Uchida, K. and Takahashi, S. and Harii, K. and Ieda, J.
             and Koshibae, W. and Ando, K. and Maekawa, S. and Saitoh, E.},
  title   = {Observation of the spin {S}eebeck effect},
  journal = {Nature},
  volume  = {455},
  pages   = {778--781},
  year    = {2008},
  note    = {\url{https://doi.org/10.1038/nature07321}}
}

@article{Uchida2010,
  author  = {Uchida, K. and Xiao, J. and Adachi, H. and Ohe, J.
             and Takahashi, S. and Ieda, J. and Ota, T. and Kajiwara, Y.
             and Umezawa, H. and Kawai, H. and Bauer, G. E. W.
             and Maekawa, S. and Saitoh, E.},
  title   = {Spin {S}eebeck insulator},
  journal = {Nat. Mater.},
  volume  = {9},
  pages   = {894--897},
  year    = {2010},
  note    = {\url{https://doi.org/10.1038/nmat2856}}
}

@article{Xiao2010,
  author  = {Xiao, J. and Bauer, G. E. W. and Uchida, K.
             and Saitoh, E. and Maekawa, S.},
  title   = {Theory of magnon-driven spin {S}eebeck effect},
  journal = {Phys. Rev. B},
  volume  = {81},
  pages   = {214418},
  year    = {2010},
  note    = {\url{https://doi.org/10.1103/PhysRevB.81.214418}}
}

@article{Yan2011,
  author  = {Yan, Peng and Wang, X. S. and Wang, X. R.},
  title   = {All-Magnonic Spin-Transfer Torque and Domain Wall Propagation},
  journal = {Phys. Rev. Lett.},
  volume  = {107},
  pages   = {177207},
  year    = {2011},
  note    = {\url{https://doi.org/10.1103/PhysRevLett.107.177207}}
}

@article{Wang2012,
  author = {Wang, X. and Guo, G. and Nie, Y. and Zhang, G. and Li, Z.},
  title   = {Domain wall motion induced by the magnonic spin current},
  journal = {Phys. Rev. B},
  volume  = {86},
  pages   = {054445},
  year    = {2012},
  note    = {\url{https://doi.org/10.1103/PhysRevB.86.054445}}
}

@article{Kovalev2012,
  author  = {Kovalev, Alexey A. and Tserkovnyak, Yaroslav},
  title   = {Thermomagnonic spin transfer and {P}eltier effects in
             insulating magnets},
  journal = {EPL},
  volume  = {97},
  pages   = {67002},
  year    = {2012},
  note    = {\url{https://doi.org/10.1209/0295-5075/97/67002}}
}

@article{Ritzmann2014,
  author  = {Ritzmann, Ulrike and Hinzke, Denise and Nowak, Ulrich},
  title   = {Propagation of thermally induced magnonic spin currents},
  journal = {Phys. Rev. B},
  volume  = {89},
  pages   = {024409},
  year    = {2014},
  note    = {\url{https://doi.org/10.1103/PhysRevB.89.024409}}
}

@article{Kong2013,
  author  = {Kong, L. Y. and Zang, J. D.},
  title   = {Dynamics of an Insulating Skyrmion under a Temperature Gradient},
  journal = {Phys. Rev. Lett.},
  volume  = {111},
  pages   = {067203},
  year    = {2013},
  note    = {\url{https://doi.org/10.1103/PhysRevLett.111.067203}}
}

@article{Chico2014,
  author  = {Chico, Jonathan and Etz, Corina and Bergqvist, Lars
             and Eriksson, Olle and Fransson, Jonas and Delin, Anna
             and Bergman, Anders},
  title   = {Thermally driven domain wall motion in {Fe} on {W}(1\,1\,0)},
  journal = {Phys. Rev. B},
  volume  = {90},
  pages   = {014434},
  year    = {2014},
  note    = {\url{https://doi.org/10.1103/PhysRevB.90.014434}}
}

@article{Schlickeiser2014,
  author  = {Schlickeiser, Frank and Ritzmann, Ulrike and Hinzke, Denise
             and Nowak, Ulrich},
  title   = {Role of Entropy in Domain Wall Motion in Thermal Gradients},
  journal = {Phys. Rev. Lett.},
  volume  = {113},
  pages   = {097201},
  year    = {2014},
  note    = {\url{https://doi.org/10.1103/PhysRevLett.113.097201}}
}

@article{Wang2014,
  author  = {Wang, X. S. and Wang, X. R.},
  title   = {Thermodynamic theory for thermal-gradient-driven domain-wall
             motion},
  journal = {Phys. Rev. B},
  volume  = {90},
  pages   = {014414},
  year    = {2014},
  note    = {\url{https://doi.org/10.1103/PhysRevB.90.014414}}
}

@article{Yan2015,
  author  = {Yan, Peng and Cao, Yunshan and Sinova, Jairo},
  title   = {Thermodynamic magnon recoil for domain wall motion},
  journal = {Phys. Rev. B},
  volume  = {92},
  pages   = {100408},
  year    = {2015},
  note    = {\url{https://doi.org/10.1103/PhysRevB.92.100408}}
}

@article{Moretti2017,
  author  = {Moretti, Simone and Raposo, Victor and Martinez, Eduardo
             and Lopez-Diaz, Luis},
  title   = {Domain wall motion by localized temperature gradients},
  journal = {Phys. Rev. B},
  volume  = {95},
  pages   = {064419},
  year    = {2017},
  note    = {\url{https://doi.org/10.1103/PhysRevB.95.064419}}
}

@article{Selzer2016,
  author  = {Selzer, Severin and Atxitia, Unai and Ritzmann, Ulrike
             and Hinzke, Denise and Nowak, Ulrich},
  title   = {Inertia-free thermally driven domain-wall motion
             in antiferromagnets},
  journal = {Phys. Rev. Lett.},
  volume  = {117},
  pages   = {107201},
  year    = {2016},
  note    = {\url{https://doi.org/10.1103/PhysRevLett.117.107201}}
}

@article{Donges2020,
  author  = {Donges, Andreas and Grimm, Niklas and Jakobs, Florian
             and Selzer, Severin and Ritzmann, Ulrike and Atxitia, Unai
             and Nowak, Ulrich},
  title   = {Unveiling domain wall dynamics of ferrimagnets in thermal
             magnon currents: competition of angular momentum transfer
             and entropic torque},
  journal = {Phys. Rev. Res.},
  volume  = {2},
  pages   = {013293},
  year    = {2020},
  note    = {\url{https://doi.org/10.1103/PhysRevResearch.2.013293}}
}

@article{islam2019t,
  author  = {Islam, M T and Wang, X S and Wang, X R},
  title   = {Thermal gradient driven domain wall dynamics},
  journal = {J. Phys.: Condens. Matter},
  volume  = {31},
  pages   = {455701},
  year    = {2019},
  note    = {\url{https://doi.org/10.1088/1361-648X/ab27d6}}
}

@article{akanda2023role,
  author  = {Akanda, M A S and Islam, M T and Wang, X R},
  title   = {Role of {SSW} on thermal-gradient induced domain-wall dynamics},
  journal = {J. Phys.: Condens. Matter},
  volume  = {35},
  pages   = {315701},
  year    = {2023},
  note    = {\url{https://doi.org/10.1088/1361-648X/accfdc}}
}

@article{islam2023role2,
  author  = {Islam, MT and Akanda, MAS and Yesmin, F and Pikul, MAJ
             and Islam, JMT},
  title   = {Role of shape anisotropy on thermal gradient-driven domain
             wall dynamics in magnetic nanowires},
  journal = {Mod. Phys. Lett. B},
  volume  = {37},
  number  = {12},
  pages   = {2350013},
  year    = {2023},
  publisher = {World Scientific},
 note    = {\url{https://doi.org/10.1142/S0217984923500136}}
  
}

@article{Pfitzner2018,
  author  = {Pfitzner, E. and Hu, X. and Schumacher, H. W. and Hoehl, A.
             and Venkateshvaran, D. and Cubukcu, M. and Liao, J.-W. and
             Auffret, S. and Heberle, J. and Wunderlich, J. and
             K{\"a}stner, B.},
  title   = {Near-field magneto-caloritronic nanoscopy on ferromagnetic
             nanostructures},
  journal = {AIP Adv.},
  volume  = {8},
  pages   = {125329},
  year    = {2018},
  note    = {\url{https://doi.org/10.1063/1.5054382}}
}

@article{OCallahan2016,
  author  = {O'Callahan, Brian T. and Raschke, Markus B.},
  title   = {Laser heating of scanning probe tips for thermal near-field
             spectroscopy and imaging},
  journal = {APL Photonics},
  volume  = {2},
  pages   = {021301},
  year    = {2016},
  note    = {\url{https://doi.org/10.1063/1.4972048}}
}

@article{Landau1935,
  author  = {Landau, L. D. and Lifshitz, E. M.},
  title   = {On the theory of the dispersion of magnetic permeability
             in ferromagnetic bodies},
  journal = {Phys. Z. Sowjetunion},
  volume  = {8},
  pages   = {153--169},
  year    = {1935},
  note    = {\url{https://doi.org/10.1016/b978-0-08-010586-4.50023-7}}
}

@article{gilbert2004,
  author  = {Gilbert, Thomas L.},
  title   = {A phenomenological theory of damping in ferromagnetic
             materials},
  journal = {IEEE Trans. Magn.},
  volume  = {40},
  pages   = {3443--3449},
  year    = {2004},
  note    = {\url{https://doi.org/10.1109/TMAG.2004.836740}}
}

@article{brown1963,
  author  = {Brown Jr., W. F.},
  title   = {Thermal Fluctuations of a Single-Domain Particle},
  journal = {Phys. Rev.},
  volume  = {130},
  pages   = {1677--1686},
  year    = {1963},
  note    = {\url{https://doi.org/10.1103/PhysRev.130.1677}}
}

@article{hinzke2008,
  author  = {Hinzke, Denise and Kazantseva, Natalia and Nowak, Ulrich
             and Mryasov, Oleg N. and Asselin, Pierre and Chantrell,
             Roy W.},
  title   = {Domain wall properties of {FePt}: From {B}loch to linear
             walls},
  journal = {Phys. Rev. B},
  volume  = {77},
  pages   = {094407},
  year    = {2008},
  note    = {\url{https://doi.org/10.1103/PhysRevB.77.094407}}
}

@article{martinez2007thermal,
  author  = {Martinez, E. and Lopez-Diaz, L. and Torres, L.
             and Tristan, C. and Alejos, O.},
  title   = {Thermal effects in domain wall motion: Micromagnetic
             simulations and analytical model},
  journal = {Phys. Rev. B},
  volume  = {75},
  pages   = {174409},
  year    = {2007},
  note    = {\url{https://doi.org/10.1103/PhysRevB.75.174409}}
}

@article{mumax,
  author  = {Vansteenkiste, Arne and Leliaert, Jonathan and Dvornik,
             Mykola and Helsen, Mathias and Garcia-Sanchez, Felipe
             and Van Waeyenberge, Bartel},
  title   = {The design and verification of {MuMax3}},
  journal = {AIP Adv.},
  volume  = {4},
  pages   = {107133},
  year    = {2014},
  note    = {\url{https://doi.org/10.1063/1.4899186}}
}

@article{islam2021thermally,
  title={Thermally assisted magnetization reversal of a magnetic nanoparticle driven by a down-chirp microwave field pulse},
  author={Islam, MT and Pikul, MAJ and Wang, XS},
  journal={J. Magn. Magn. Mater.},
  volume={537},
  pages={168174},
  year={2021},
publisher={Elsevier},
  note    = {\url{https://doi.org/10.1016/j.jmmm.2021.168174}}
}

@article{Markus2022,
  title = {Walker breakdown of {B}rownian domain wall dynamics},
  author = {Wei\ss{}enhofer, Markus and Selzer, Severin and Nowak, Ulrich},
  journal = {Phys. Rev. B},
  volume = {106},
  issue = {10},
  pages = {104428},
  numpages = {11},
  year = {2022},
  month = {Sep},
  publisher = {American Physical Society},
  note    = {\url{https://doi.org/10.1103/PhysRevB.106.104428}}
}

@article{Nakatani2003,
  author  = {Nakatani, Yoshinobu and Thiaville, Andr{\'e} and Miltat,
             Jacques},
  title   = {Faster magnetic walls in rough wires},
  journal = {Nat. Mater.},
  volume  = {2},
  pages   = {521--523},
  year    = {2003},
  note    = {\url{https://doi.org/10.1038/nmat931}}
}

@article{Mochizuki2014,
  author  = {Mochizuki, M. and Yu, X. Z. and Seki, S. and Kanazawa, N.
             and Koshibae, W. and Zang, J. and Mostovoy, M.
             and Tokura, Y. and Nagaosa, N.},
  title   = {Thermally driven ratchet motion of a skyrmion microcrystal
             and topological magnon {H}all effect},
  journal = {Nat. Mater.},
  volume  = {13},
  pages   = {241--246},
  year    = {2014},
  note    = {\url{https://doi.org/10.1038/nmat3862}}
}
\end{document}